# Self-consistent particle-in-cell simulations of fundamental and harmonic plasma radio emission mechanisms.

J. O. Thurgood[1] and D. Tsiklauri

School of Physics and Astronomy, Queen Mary University of London, Mile End Road, London, E1 4NS, United Kingdom
e-mail: j.thurgood@qmul.ac.uk

**ABSTRACT**

*Aims.* The simulation of three-wave interaction based plasma emission, thought to be the underlying mechanism for Type III solar radio bursts, is a challenging task requiring fully-kinetic, multi-dimensional models. This paper aims to resolve a contradiction in past attempts, whereby some studies indicate that no such processes occur.

*Methods.* We self-consistently simulate three-waved based plasma emission through all stages by using 2D, fully kinetic, electromagnetic particle-in-cell simulations of relaxing electron beams using the EPOCH2D code.

*Results.* Here we present the results of two simulations; Run 1 ($n_b/n_0 = 0.0057$, $v_b/\Delta v_b = v_b/V_e = 16$) and Run 2 ($n_b/n_0 = 0.05$, $v_b/\Delta v_b = v_b/V_e = 8$), which we find to permit and prohibit plasma emission respectively. We show that the possibility of plasma emission is contingent upon the frequency of the initial electrostatic waves generated by the bump-in-tail instability, and that these waves may be prohibited from participating in the necessary three-wave interactions due to frequency conservation requirements. In resolving this apparent contradiction through a comprehensive analysis, in this paper we present the first self-consistent demonstration of fundamental and harmonic plasma emission from a single-beam system via fully kinetic numerical simulation. We caution against simulating astrophysical radio bursts using unrealistically dense beams (a common approach which reduces run time), as the resulting non-Langmiur characteristics of the initial wave modes significantly suppresses emission. Comparison of our results also indicates that, contrary to the suggestions of previous authors, an alternative plasma emission mechanism based on two counter-propagating beams is unnecessary in an astrophysical context. Finally, we also consider the action of the Weibel instability which generates an electromagnetic beam mode. As this provides a stronger contribution to electromagnetic energy than the emission, we stress that evidence of plasma emission in simulations must disentangle the two contributions and not simply interpret changes in total electromagnetic energy as evidence of plasma emission.

## 1. Introduction

The emission of electromagnetic radiation at the local electron plasma frequency and twice the plasma frequency, otherwise known as *fundamental* and *harmonic plasma emission*, has been observed to be a common phenomenon in both astrophysical and laboratory plasmas. Astrophysical examples include Type II & III solar radio bursts (Lin et al. 1981, 1986; Dulk et al. 1984; Goldman 1983; Robinson et al. 1993a,b, 1994; Reid & Ratcliffe 2014), outer heliospheric radio emission (Kurth et al. 1984), and in planetary electron foreshocks (Etcheto & Faucheux 1984; Moses et al. 1984; Fuselier et al. 1985; Lacombe et al. 1985). Related processes in laboratory plasmas have also been discussed by, e.g. Hutchinson et al. (1978); Benford et al. (1980); Whelan & Stenzel (1981); Intrator et al. (1984); Whelan & Stenzel (1985). Furthermore, radio bursts of millisecond duration were recently discovered in the 1 GHz band Loeb et al. (2014), with strong evidence that they come from $\sim$ 1 Gpc distances, implying extraordinarily high-brightness temperature. Lyubarsky (2014) proposed that these bursts could be attributed to synchrotron maser emission from relativistic, magnetized shocks. Since fast radio bursts are associated with flaring stars (Loeb et al. 2014) it may well be that the plasma emission caused by an electron beam (as opposed to shocks) may be a viable mechanism.

A long-standing, multiple-stage model for the fundamental and harmonic emission based on subsequent nonlinear three-wave interactions, has been considered extensively by authors including, but not limited to, e.g., Ginzburg & Zhelezniakov (1958); Melrose (1980, 1987); Cairns (1987); Robinson et al. (1994). In the first stage of this model, electron beams which are excited in the low-corona are susceptible to the bump-in-tail instability as they propagate through the background plasma, and thus the beams can generate Langmuir waves ($L$). The Langmuir waves can then decay into ion-sound waves ($S$) and electromagnetic waves at the local plasma frequency ($T_1$) in the process $L \rightarrow S + T_1$, which is thought to explain fundamental emission. Harmonic emission occurs due to two subsequent three-wave interactions; firstly, the electrostatic decay of forward-propagating Langmuir waves into ion-sound and backward-propagating Langmuir waves ($L'$) $L \rightarrow L' + S$ (or equivalently, backscattering off ion-sound waves $L + S \rightarrow L'$), and secondly, the coalescence of counter-propagating Langmuir waves to produce electromagnetic waves at twice the local plasma frequency ($T_2$), $L + L' \rightarrow T_2$.

Whilst many numerical works have considered individual stages of these fundamental and harmonic emission mechanisms, only a few attempts have been made to self-consistently verify the processes using fully-kinetic numerical simulations. The primary reason for this is that the self-consistent simula-tion of the emission process is necessarily a computationally demanding task that requires a full electromagnetic treatment in two dimensions which must be sufficiently large to contain all participating wavelengths. Two dimensions are required because the emission formulas (three-wave interaction probabilities) for both $L \rightarrow S + T_1$ and $L + L' \rightarrow T_2$ are only non-zero when a vector-product exists between participating wave vectors, see e.g., Eqs. (26.24) and (26.25) from Melrose & McPhedran (2005). As the instability time scales with the inverse of the beam-to-background density ratio (e.g., the quasilinear relaxation timescale is $\tau_{ql} = (n_0/n_b)\,(v_b/\Delta v_b)^2\,\omega_{pe}^{-1}$), direct simulation of astrophysical parameter regimes where beams are typically diffuse ($n_b/n_0 \approx 10^{-5} - 10^{-6}$) requires the simulation to run for tens-to-hundreds of thousands of electron plasma periods. Furthermore, this is compounded in that huge particle numbers are required to correctly resolve the expected quasilinear relaxation timescales with the particle-in-cell (PIC) method (Ratcliffe et al. 2014; Lotov et al. 2015). Lotov et al. (2015) found empirically that for beam-plasma systems the number of pseudoparticles per cell required scales as the inverse of the fraction of (real) particles which are in resonance with the beam. In particular, for diffuse astrophysical beams, this leads to a very high pseudoparticle requirement of tens-to-hundreds of thousands particles per cell.

In the handful of papers that have considered the fully self-consistent problem with the PIC method, a comparative review suggests that the previous results are somewhat contradictory. In the most extreme, some authors interpret the results of their experiments as evidence supporting the three-wave based mechanisms (Kasaba et al. 2001; Umeda 2010), whereas others report that the such processes do not proceed in their simulations, and rather must invoke different mechanisms to produce emission; such as the requirement for a counter-propagating beam (Ganse et al. 2012b,a, 2014; Timofeev & Annenkov 2014), or linear mode conversion off density inhomogeneities as per Sakai et al. (2005). Although, note that in the latter work they did not present results for a homogeneous setup, so arguably the underlying process responsible for emission in their work is not conclusively demonstrated.

More subtly, further ambiguity exists in the details of the simulations proporting to show evidence of plasma emission. Firstly, limited attention is paid towards distinguishing the peaks in spectral power at the expected frequencies, which are taken as evidence of emission occurring, from thermal and noise levels. This is of crucial importance as noise organising itself as electromagnetic waves (most obviously manifest as power enhancements along dispersion curves in $(\omega, k)$-diagrams) is a typical and unavoidable feature of PIC simulations. Secondly, pseudoparticle numbers used are typically low (e.g., Kasaba et al. (2001) uses 16 particles per cell for background partxicles and 4 for the beam particles, and Umeda (2010) uses 256 particles per cell per species), and few papers report convergence testing of results. Umeda (2010) reported that increasing pseudoparticle numbers diminished the signal associated with harmonic emission. Under-resolved particle numbers could contribute to excessive noise on the EM dispersion curves (and be mis-interpreted as emission as per the first point) and also result in poor replication of physical time-scales as per Ratcliffe et al. (2014) and Lotov et al. (2015). Finally, all of the previous works consider systems with high beam-to-background density ratios (all take $n_b/n_0 > 1\%$) and are so in the *strong beam* regime. As astrophysical systems typically have diffuse electron beams, this is presumably motivated by the faster relaxation times for dense beams allowing for significantly reduced computational requirements. However, substantial theoretical work has repeatedly shown that beam-plasma systems are expected to posses incredibly sensitive parameter spaces. In particular, dense beams are associated with strong modification of the dispersion relationships (O'Neil & Malmberg 1968; Cairns 1989), and so we raise the question *how valid are strong beam simulations of a process that is analytically (in quasilinear theory) prescribed for weak beams?*

**Table 1.** Simulation Parameters

| Run | $n_b/n_0$ | $v_b/\Delta v_b = v_b/V_e$ | Notes |
|---|---|---|---|
| 1 | 0.0057 | 16 | Weak beam |
| 2 | 0.05 | 8 | Dense beam |
| 3 | 0.05 | 8 | Two counter-propagating beams |

The primary aim of this paper is to carry out detailed 2D PIC simulations of varying beam-plasma systems to resolve the apparent contradictions discussed above; in other words to clarify whether the three-wave interaction based, single-beam plasma emission can be self-consistently shown to proceed as expected for astrophysical parameter regimes, or whether the two-beam mechanism is required as suggested by Ganse et al. (2012b). In resolving this apparent contradiction through a comprehensive analysis, in this paper we present the first self-consistent demonstration of fundamental and harmonic plasma emission from a single-beam system via fully kinetic numerical simulation. The paper is structured as follows; in Section 2 we detail the numerical methods used and the initial conditions/setup for the experiments, in Section 3 we present the results, in Section 4 we discuss their implications and in Section 5 we draw concluding remarks.

## 2. Numerical Setup

The simulations are carried out using EPOCH2D, a 2.5D Birdsall and Langdon type PIC code with a core solver based on the PSC code of Rhul (Chapter 2 of Bonitz & Semkat (2006); see also Brady & Arber (2011)). In this paper we present the results of three numerical experiments. The first two simulations vary in beam-to-background number density ratio, beam temperature and speed as $n_b/n_0 = 0.0057$, $v_b/\Delta v_b = v_b/V_e = 16$ (*Run 1*) and $n_b/n_0 = 0.05$, $v_b/\Delta v_b = v_b/V_e = 8$ (*Run 2*). Thus, Run 1 is weaker beam than considered in previous PIC simulations of plasma emission (but still orders of magnitude larger than typical astrophysical plasmas) and Run 2 is more comparable to the past simulations. The initial conditions are chosen for two reasons; firstly, their 1D analogues have recently been considered by Baumgärtel (2014), who noted a difference in the nature of the electrostatic decay processes in the two regimes, and so we may explore the consequences for emission processes in our extension to 2D. Secondly, each set-up gives comparable parameters

$$P = (n_b/n_0)^{1/3}(v_b/\Delta v_b) \tag{1}$$

of $P = 2.85$ and $P = 2.94$, which is a measure of the kinetic versus fluid nature of the instability. As per the parameter study of Cairns (1989), these $P$ values indicate a predominantly kinetic instability (NB: a key finding of that paper was that the kinetic criterion $P < 1$ sometimes reported by other authors is not accurate). These experiments are presented in order to illustrate the sensitivity to different beam parameters, and to investigate the consequences for efficiency of plasma emission. Furthermore, a

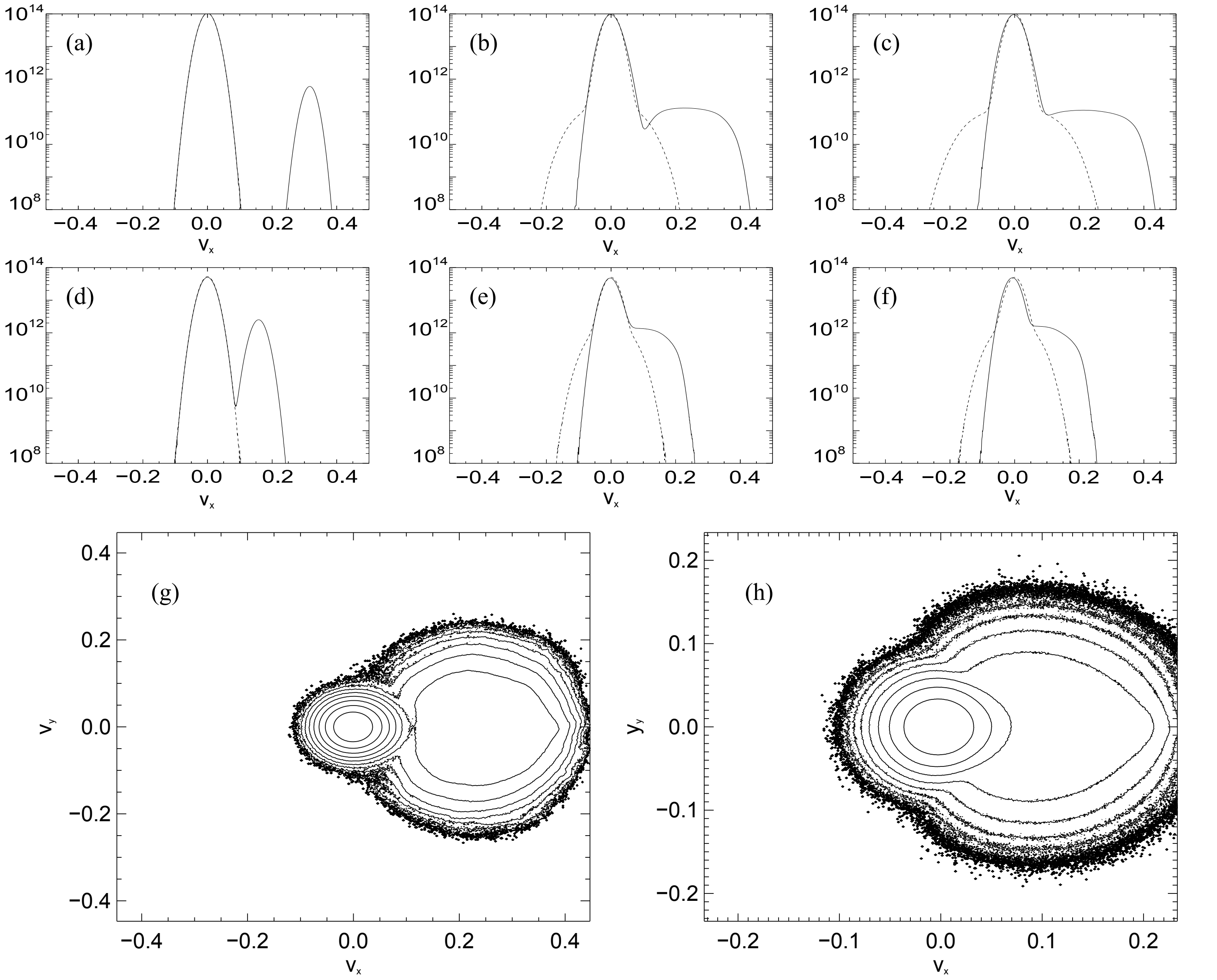

**Fig. 1.** Electron velocity distribution functions for Run 1 (a-c) and Run 2 (d-f) at times $t\omega_{pe}^{-1}$ = 0, 300, 600 respectively. The solid line shows the velocity distribution aligned with the beam ($v_x$) and the dashed line shows the transverse velocity ($v_y$). Panels (g) and (h) are representative of the behaviour in 2D phase-space for Run 1 and 2 respectively at $t\omega_{pe}^{-1} = 300$. Note the plateau formation in $v_x$ with the saturation of the bump-in-tail instability, and the transverse heating of the beam electrons due to the Weibel instability.

third experiment (*Run 3*) extends Run 2 by considering the setup in the presence of an additional counter-streaming (but otherwise identical) electron beam in order to investigate the differences between single-beam and two-beam plasma emission, and the circumstances in which a two-beam emission mechanism may be necessary. The difference in parameters between the three runs is summarised in Table 1

All three experiments share common background plasma parameters, which are chosen as follows; the background number density $n_0 = n_e = n_i = 4 \times 10^6$ m$^{-3}$, background temperature $T_e = 200$ eV, $T_i = 0.73T_e$ and the mass ratio $m_i/m_e = 1836$. The background magnetic field is set to zero (i.e., we do not consider a weak guide field aligned with the beam). The simulations are ran for $1000\,\omega_{pe}^{-1}$ in a domain of size $L_x = L_y = 600\lambda_D$, where the cell width is equal to the Debye length $\Delta_x = \Delta_y = \lambda_D$, and periodic boundary conditions are applied. For each simulation, convergence testing was carried out for pseudoparticle counts of 125, 250, 500 and 1000 particles per cell per species (PPCPS). The convergence tests consisted of (1) ensuring convergence of relaxation time of the bump-in-tail instability (i.e., ensuring convergence for figures similar to Fig. 1) and (2) ensuring convergence of spectral energy density levels associated with different wave modes to within an order of magnitude (NB: convergence to specific numbers is impossible due to the random nature of the PIC code; increasing the particle count improves statistical representation). The convergence tests found tolerable convergence of the results for 500 and 1000 PPCPS. The results presented here are for 1000 PPCPS, which required ~ 72 hours of computing time each on 480 2.50GHz Intel Xeon cores.

## 3. Results

First we consider the evolution of the electron population in velocity phase space. Figure 1 shows electron velocity distribution functions for Run 1 (a-c) and Run 2 (d-f) at times $t\omega_{pe}$ = 0, 300, 600 respectively, where the solid line shows the veloc-ity distribution aligned with the beam ($v_x$) and the dashed line shows the velocity distribution across the beam ($v_y$). In $f(v_x)$ phase space (solid lines), we see the saturation of the initial bump-in-tail instability and the merging of the beam electrons to the bulk distribution, giving the characteristic plateau formation. This is observed to occur by around $t \approx 60\,\omega_{pe}^{-1}$ (Run 1) and $t \approx 600\,\omega_{pe}^{-1}$ (Run 2), which are in excess of time-scales predicted by the quasi-linear formula. We however, confirmed that this is a realistic relaxation time in our convergence testing, rather than being due to poor particle counts as reported by Ratcliffe et al. (2014). The discrepancy is due to the use of the relatively strong beams (in terms of kinetic energy); the applicability of the weak turbulence/quasi-linear results require that the ratio of beam to background kinetic energies be much less than unity. In $f(v_y)$ phase space (dashed lines) we see the heating of the electrons in both runs, due to the action of the Weibel instability. The Weibel instability is expected to cause perpendicular heating in beam-plasma systems except in the case of strong magnetic fields or in the reduction to a 1D system, as discussed in detail by Karlický (2009). Again, this occurs on a faster timescale and saturates faster in Run 2 compared to Run 1. Panels (g) and (h) of Figure 1 show visualisations of the above behaviour in 2D phase-space at a representative time($t\omega_{pe}^{-1}$ = 300) for Run 1 and Run 2 respectively. These panels highlight that it is the beam population, rather than the background electrons, which are susceptible to heating in the $y$-direction, which has consequences that we will explore later. Overall, for both runs, we see the relaxation of electron phase space to asymptotic states during the course of the experiment.

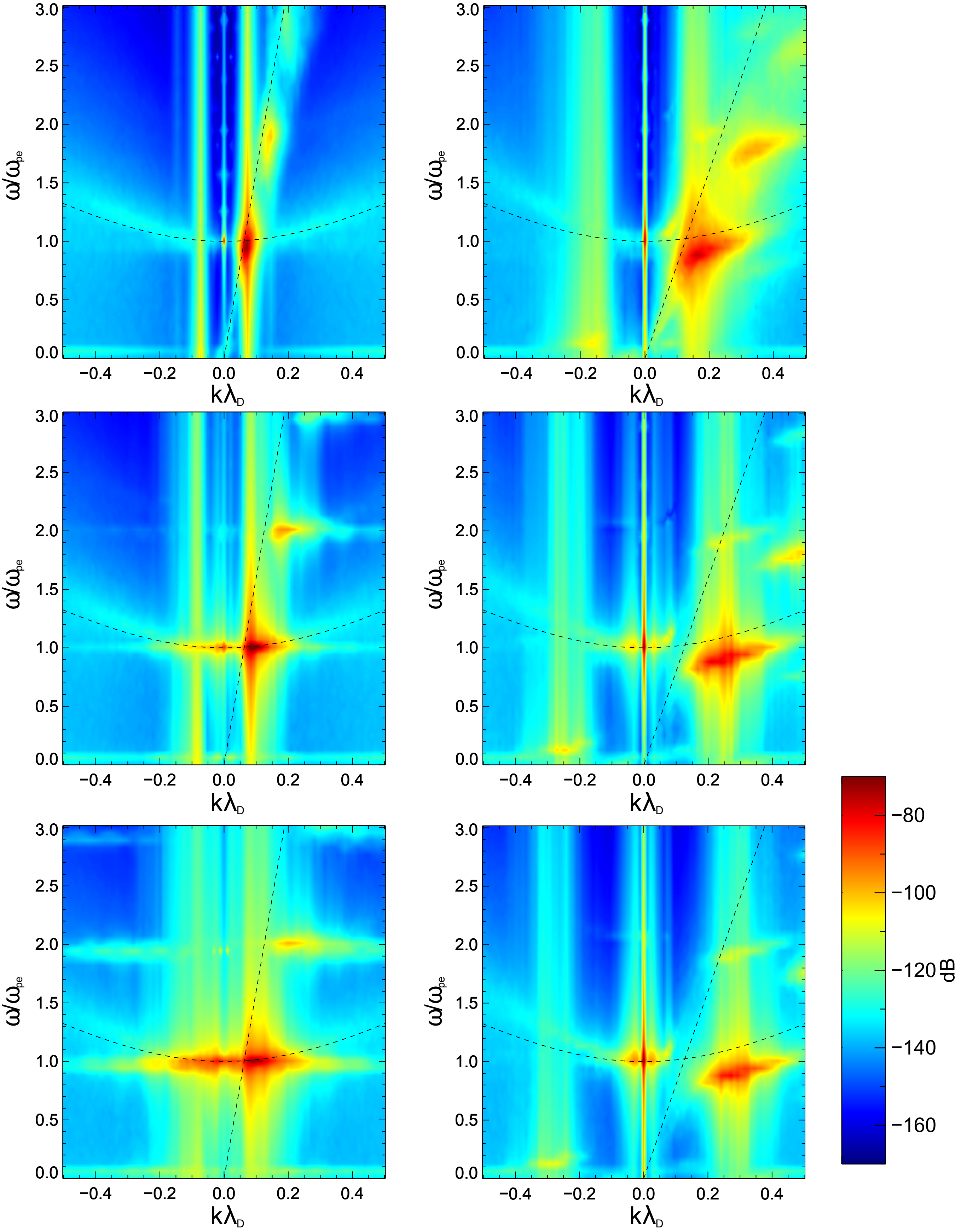

**Fig. 2.** Fast fourier transforms of $E_x$ for Run 1 (left) and Run 2 (right) over times $50 < t\omega_{pe}^{-1} < 150$, $450 < t\omega_{pe}^{-1} < 550$, $850 < t\omega_{pe}^{-1} < 950$, from top to bottom respectively.

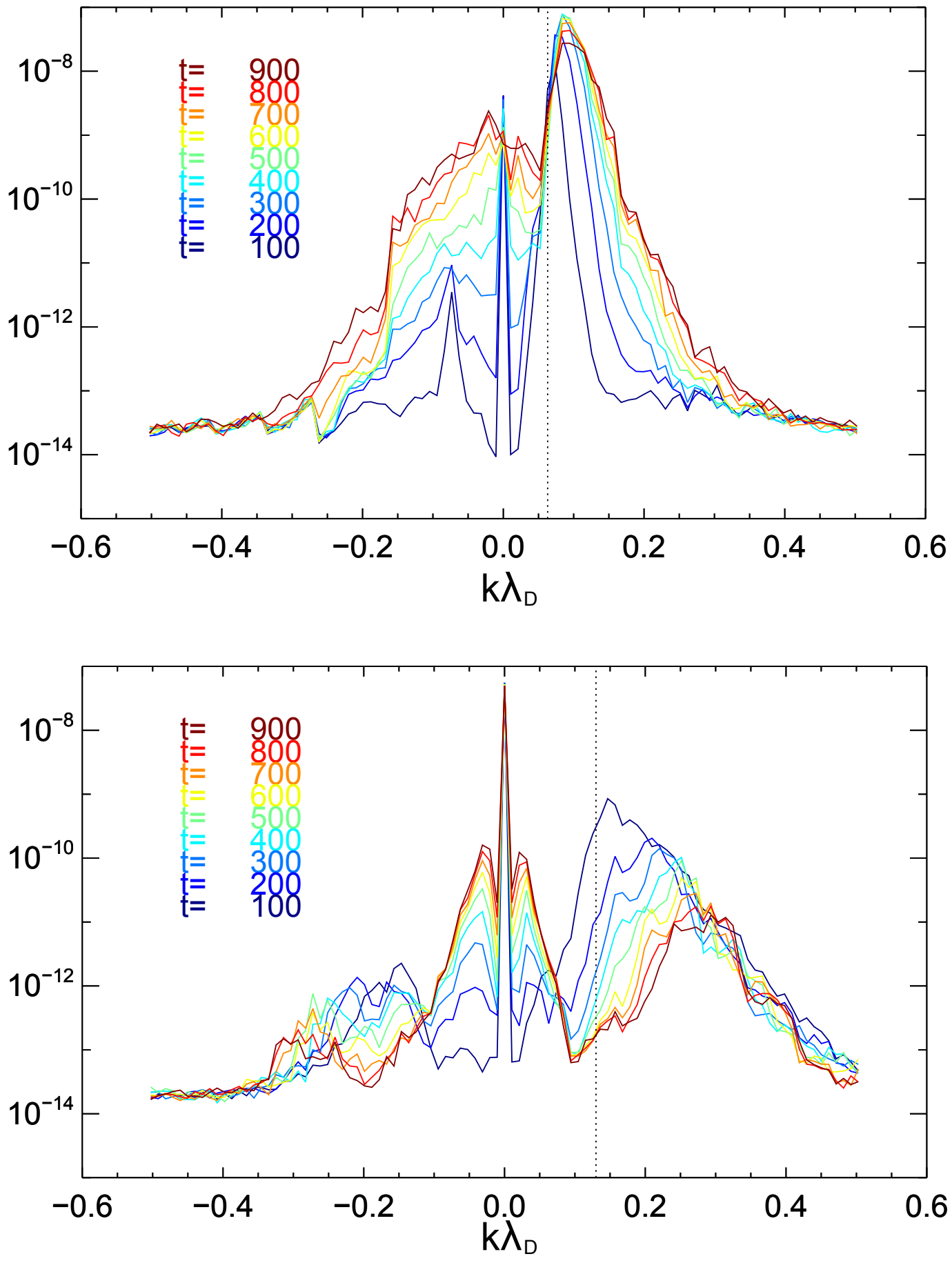

**Fig. 3.** Spectral energy density (J m$^{-3}$) along the Langmuir wave dispersion curve for Run 1 (top) and Run 2 (bottom) for subsequent $100\,\omega_{pe}^{-1}$ time windows (centred $\pm 50\,\omega_{pe}^{-1}$ about the indicated time). Note the development of a broad spectrum in negative $k_x$ in Run 1 with energy densities several orders of magntude larger than the negative $k_x$ spectrum of Run 2. The dotted lines show the expected initial $k_x$ value of the initially growing electrostatic modes.

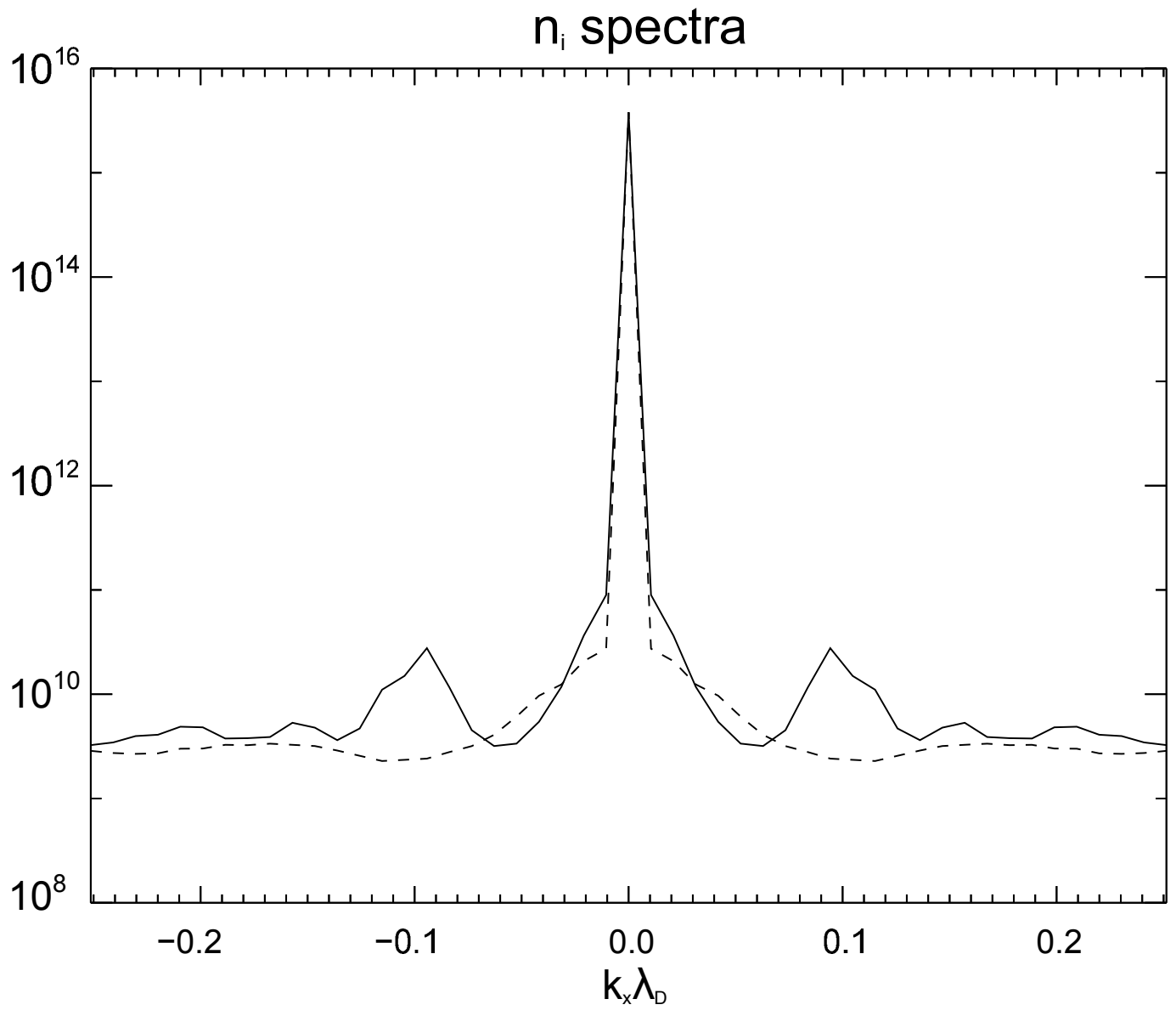

**Fig. 4.** $k_x$ Spectra of ion density fluctuations (integrated in $y$ and $t$) for Run 1 (solid line) and Run 2 (dashed line). Note the presence of peaks at $|k| \approx 2k_L$, where $k_L$ is the predominant wavenumber associated with the growing Langmuir waves found in in Run 1, suggestive of the action of Langmuir decay / scattering processes involving ion-acoustic waves. No analogous peak for Run 2 is found.

We now turn our attention to electrostatic waves generated by the beam relaxation. Figure 2 shows the evolution of the parallel electric field energy in $(k_x, \omega)$ space by considering 2D, windowed Fourier transforms in $(x, t)$ space over subsequent 100 $\omega_{pe}^{-1}$ periods, which are then integrated over the $y$ direction. The time window size was determined by experimentation and provides the best balance between frequency resolution ($\Delta\omega$) and time cadence, allowing us to track the evolution in both Fourier space and time, whilst preserving sufficient frequency resolution to reasonably compare spectra to expected dispersion curves. The left-hand column shows the spectral electrical energy density for Run 1, and the right for Run 2, over time periods $50 < t\omega_{pe} < 150$, $450 < t\omega_{pe} < 550$ and $850 < t\omega_{pe} < 950$ (top to bottom). The dispersion curves for the expected beam modes ($\omega/k = v_b$) and (unmodified) Langmuir modes ($\omega^2 = \omega_{pe}^2 + 3k^2V_e^2$) are overlaid.

For Run 1, we find that the majority of initial growth occurs on the intersection of the beam-mode and Langmuir mode. Over time, the power is efficiently transferred to negative $k_x$ via decay/backscattering processes of the type $L \rightarrow L' + S$ , thus creating a seed population of forward and backward propagating waves which are possible candidates for participation in the coalescence $L + L' \rightarrow T_2$ required for harmonic emission. The backwards (negative $k_x$) portion of the Langmuir wave dispersion curve appears to be slightly shifted towards higher $\omega$ at larger $k_x$ than expected in a quiescent plasma, which is consistent with the solutions of Cairns (1989), who numerically determined the changes to plasma dispersion relation due to the presence of a beam. We can more closely inspect and compare the energy evolution associated with electrostatic waves by plotting the evolution of energy density along the curve $\omega^2 = \omega_{pe}^2 + 3k^2V_e^2$, as shown in Fig. 3, which illustrates the previous discussion. Note that, in Fig. 3, we have sampled the frequency range $\pm\Delta\omega$ above and below the curve to account for observed deviations in $\omega$ from the unmodified Langmuir wave. The role of three-wave backscatter/decay progress of the type $L \rightarrow L' + S$ in the creation of the counter-propagating Langmuir wave propagation is confirmed by the presence of enhanced ion-density fluctuations at the expected $k_S$ ($k_S \approx 2k_L$, as shown in Figure 4). Further, this is consistent with the PIC simulations of Baumgärtel (2014) who considered the 1D analogue of this system (i.e., the same beam-plasma setup as Run 1 in a 1D PIC code) which afforded superior $k_x$ resolution by using box sizes which are inaccessible to our 2D simulations, unambiguously confirming the action of decay/backscattering processes in the the system's 1D counterpart.

Thus, for Run 1, we find qualitative evolution of the power spectra that is consistent with the first two stages of the single-beam, three-wave based harmonic emission mechanism, namely (1) the efficient coupling of the beam to electrostatic waves, and subsequently (2) the generation of an eligible seed population of forward and backward propagating electrostatic waves via nonlinear three-wave interactions. Quantitatively, the seed population grows to be at least 2-3 orders of magnitude in excess of background levels by $t\omega_{pe} = 900$. Additional features can also be identified, most notably the so-called high-harmonic electrostatic nonlinear plasma waves (Rhee et al. 2009; Yi et al. 2007; Yoon et al. 2003). As pointed out by Yoon et al. (2003) these modes are nonlinear eignemodes that exist by virtue of the nonlinear plasma response to the presence of the beam, and do not exist as 'natural' modes in quiescent plasmas. We also note the presence of a spectral peak at $k_x = 0$. This corresponds to a beam-aligned, standing mode of the electric field ($E_x$) oscillating at the local plasma frequency, which is present from the simulation initialisation (thus, before the instability onset). It is caused by the non-zero initial current imposed by the beam at $t = 0$, and has been discussed in detail by Baumgärtel (2013). Whilst it is possible to remove this mode by introducing compensating drift velocity to the background electrons, we tolerate its presence as it is unclear whether such a compensation is physically appropriate (in particular, it may influence the correct return-current processes). Regardless, we have found that the amplitude associated with this mode, *in all runs* was much less than that of the Langmuir waves generated after the instability onset and as such it does not go on to effect the dynamics of the Langmuir waves which participate in the plasma emission mechanism. The influence and relative amplitude of the beam-aligned mode is most easily determined from time-distance diagrams (not shown here, for brevity), however its minor influence on the energy budget can be seen Figure 8, which is discussed in more detail later in this section.

However, in Run 2 we do not find analogous behaviour. The beam mode 'decouples' from the expected Langmuir / electrostatic mode and the main concentration of power associated with the beam mode drifts towards higher $k_x$ as time evolves (Fig. 2). Furthermore, it is primarily concentrated below the frequency expected of forward-propagating electrostatic waves, and so there exists the possibility of three-wave interactions becoming strongly limited by frequency conservation requirements as shown by Cairns (1989); this is discussed further in Section 4. Growth on both the forward and backward portions of the Langmuir curve itself (Fig. 3) is predominantly at small $|k_x|$. At all times, the backward (negative $k_x$) spectrum is at least an order of magnitude weaker than possible counterparts in Run 1 (but generally ranges from being $1 - 3$ orders weaker), supporting the conjecture that the decay process has been inhibited in the case of Run 2. Additionally, an absence of significant ion-density power enhancement in Run 4 is consistent with an absence of decay/scattering processes in excess of noise levels, contrary to Run 1 (cf. Fig 4).

We now turn our attention to the consequences of the notably different nature of electrostatic wave growth and evolution in Runs 1 and 2 for plasma emission by considering equivalent figures for the transverse magnetic field $B_z$. In Figure 5 we see the evolution of spectral energy density in $B_z$ in Run 1 (left) and Run 2 (right) over the same timesteps considered in Figure 2. For both runs we see the growth of power along the electromagnetic dispersion curves (overplotted), some of which is associated with the PIC noise discussed in Section 1. When investigating plasma emission processes with the PIC method, it is crucial that any signals of emission at the expected wavenumbers and frequencies is clearly distinguished from such noise. For the results presented here, we estimate the noise threshold from the average power on the curves away from the frequencies that are expected to be enhanced by plasma emission mechanisms ($\omega = \omega_{pe}$, and its harmonics), and find it to be of the order $\sim -140\,\mathrm{dB} = 10^{-14}\,\mathrm{J\,m^{-3}}$. Hence forth, we only consider signal on electromagnetic dispersion curves that is distinguished above this threshold as an indicator of plasma emission occurring.

With regards to indicators of fundamental emission in Figure 5, in both experiments we see an enhancement of power at $\omega = \omega_{pe}$ for small $k_x$. We can take a closer look at the enhancement in Figure 6, which considers the distribution of spectral energy density over $k_x$ at $\omega = \omega_{pe}$ (similar to Figure 3). At small $k_x$ for Run 1, we observe the growth of a peak in excess of the noise levels, which reaches $\sim 10^{-12}\,\mathrm{J\,m^{-3}}$, two orders of magnitude above the threshold, in the time window $900 < t\omega_{pe} < 1000$. The peak's apex is slightly shifted towards negative $k_x$, located at $k_x\lambda_D = -0.01$ . Note that, the expected wavenumber from the fundamental emission process $L \rightarrow S + T_1$ for Run 1's beam parameters is $k_{T1}\lambda_D \approx 0.002$, below our sampling rate of $\Delta_k$. As such, confirming whether the signal is predominantly concentrated at the expected wavenumber is not possible for these simulations and would require larger spatial domains. For Run 2, we see the development of a weaker, narrower signal peaked at $k_x = 0$ which is at most an order of magnitude above the noise threshold, i.e., its peak spectral energy density is one order of magnitude below the equivalent signal in Run 1. Thus, for both runs we see a signal that is consistent with fundamental emission, but it is comparatively weaker for Run 2. Additionally, for both runs, in Figures 5-7 we see features which are not associated with electromagnetic radiation (i.e, the peak at larger positive $k_x$ in figure 6). These are driven by the perpendicular heating of the beam electrons (Weibel instability) as seen in Figure 1 and are manifest its various harmonics (cf. Figures 2 and 5). These *electromagnetic beam modes* are found in the $E_y$ spectrum (which is not shown here, but is qualitatively as per Fig. 5). On close inspection, these modes identified can be seen in other authors works (eg., Plate 2 of Kasaba et al. (2001)) although they recieved no attention at the time, presumably due to negligence in considering perpendicular phase space, and so the realisation of the role of the Weibel instability, which was only first demonstrated by Karlický (2009). The tenancy for the Weibel instability to drive non-maxwellian, electromagnetic eignemodes at $\omega = kv_b$ was discovered experimentally by Urrutia & Stenzel (1984) and explored theoretically by Goldman & Newman (1987). To our knowledge, this is the first time the the electromagnetic beam mode has been discussed in the context of plasma emission. Relatively large amounts of power are concentrated in these modes (more than that emitted by the plasma emission) and so we caution that total power contained in the electromagnetic field is not exclusively associated with the plasma radio emission and must not be interpreted as such. This will be important in our later consideration of the overall energy budget of the systems.

In Figure 5, as time advances we see the development of a clear enhancement near $\omega = 2\omega_{pe}$ on the electromagnetic dispersion curve for Run 1, but equivalent signal for Run 2 can only be poorly distinguished against the EM noise levels. Again, this is examined more closely in Figure 7 by considering the spectra at fixed $\omega = 2\omega_{pe}$. For Run 1, we find four clearly distinguished spectral peaks close to the expected $|k_x|$ values for harmonic emission generated by the coalescence $L + L' \rightarrow T_2$ which grow to be in excess of two orders of magnitude above noise levels ($\approx 10^{-12}\,\mathrm{J\,m^{-3}}$) by the time period $900 < t\omega_{pe} < 1000$. For Run 2, however, we only see the slow development of small peaks that are comparable with the noise, with a maximum spectral energy density of $\approx 10^{-14}\,\mathrm{J\,m^{-3}}$. Thus, signals consistent with harmonic plasma emission that are clearly distinguished from background levels are detectable in only the case of Run 1. Note that the reason that we see four peaks in $k_x$ at $\omega = 2\omega_{pe}$, rather than two, is simply an artefact of projecting obliquely propagation emission into $k_x$ space. If we instead consider $B_z(k_x, k_y)$ at fixed $\omega = 2\omega_{pe}$, we find power enhancements suggesting emission at angles (relative to beam direction) of 22°, 68°, 112°, and 158° which is in agreement with that reported by Umeda (2010).

In Figure 8 we consider the energy evolution of the system, normalised to the initial beam kinetic energy. For the energy associated with the parallel electrostatic component ($\int 0.5\varepsilon_0 E_x^2\,\mathrm{dV}$), for both runs, we initially see the 'beam-aligned mode' ($k_x = 0$ mode). After the instability onset, the energy associated with $E_x$ grows to around 10% of the beam kinetic energy in both cases. Thus, the driven electrostatic energy rapidly overpowers the $k_x = 0$ mode, as discussed earlier. This is accompanied by an energy increase associated with $E_y$ ($\int 0.5\varepsilon_0 E_y^2\,\mathrm{dV}$), which is predominantly by the perpendicular acceleration of the beam particles by the Weibel instability, but will also contain the electric energy associated with any plasma emission. The energy associated with the corresponding magnetic field component, $B_z$ ($\int 0.5\mu_0^{-1} B_z^2\,\mathrm{dV}$), similarly grows after the instability onset, then decays, and subsequently begins to grow again at later times in both runs. Here we stress that the *initial growth* is associated with the direct generation of *electromagnetic beam modes* by the Weibel instability. The later growth is associated with the comparatively weak fundamental and harmonic emission processes identified in earlier paragraphs. Here, we stress caution against interpreting any increase in total electromagnetic energy as evidence of emission as has been the case in some previous studies. For example, from Figure 8, one may be tempted to conclude a similar efficiency and timescale of presumed 'emission processes' for both runs, although as we have seen in our earlier analysis that emission, particularly harmonic, is much weaker or non-existent in Run 2. It is crucial to appreciate that the electromagnetic beam mode is the predominant contributor to total electromagnetic energy, and so emission energy and efficiency is better identified by the spectral methods we have employed which are able to disentangle the contributions of different modes. We also note that energy in $B_x$, $B_y$ and $E_z$ does not grow beyond noise levels and as such contain only a negligible portion of the systems energy. From this we may infer that emission is linearly polarised, as it is only manifest in $B_z$ and not $B_y$ (equivalently, manifest in $E_y$ and not $E_z$). This is to be expected, due to the absence of an initial magnetic field. Plasma emission from Type III solar radio bursts has be observed to be typically weakly circularly polarised (McLean 1971; Suzuki & Sheridan 1977; Dulk & Suzuki 1980; Suzuki & Dulk 1985), with harmonic emission having a lower degree of polarisation than fundamental. Finally, we note that when a total system energy is calculated we find that it is well-conserved, to an accuracy of 0.03% during the simulation lifetime.

## 4. Discussion

We have found the following key features in the comparison of the wave dynamics and consequences for plasma emission in two different (single) beam-plasma systems;

1. The growth of the beam mode on the Langmuir wave dispersion curve in Run 1, and apparent 'decoupling' of the beam and Langmuir modes in Run 2, whereby the presence of the denser beam has significantly modified the nature of the forward propagating electrostatic wave modes.
2. The apparent difference in susceptibility of the two systems to processes of the type $L \rightarrow L' + S$, with consequences for the production of a seed population of counter-propagating Langmuir/Electrostatic waves.
3. We observe fundamental emission distinguishable above the noise threshold in both cases. At its peak, the spectral energy density associated with the emission is two-orders of magnitude stronger than the noise threshold in Run 1, but only one-order of magnitude greater in Run 2. Thus, the fundamental emission is weaker in the strong beam case of Run 2.
4. We observe that harmonic emission is clearly distinguishable above the noise level for Run 1 (two orders of magnitude at peaks), but only weak signals comparable to background levels are detected in Run 2. Thus, harmonic emission is much less effective in the case of Run 2, possibly to the point of being prohibited altogether.

That statement (4) is the case, namely that harmonic emission is only observed at twice the background level in Run 2, and that it is in its most generous interpenetration weak in Run 2, is a direct result of (2). Run 1 has been demonstrated to posses a larger seed population of forward and backward propagating electrostatic waves for participation in the processes of the type $L + L' \rightarrow T_2$. Thus, as is it may be expected, for harmonic emission the efficiency of electrostatic decay processes $L \rightarrow L' + S$ is of primary importance. The fundamental difference in the efficiency of the transfer of energy from the beam mode to a population of forward and backward propagating Langmuir waves can be explained in terms of (1), i.e. the nature of the generated forward propagating electrostatic waves in the two parameter regimes. Crucially, in Run 2 the forward propagating electrostatic mode associated with the beam is primarily concentrated below the plasma frequency (cf. Fig. 2). Reading from Figure 2, the packet is concentrated at approximately $0.9\,\omega_{pe}$. To participate in the three wave interactions $L \rightarrow L' \pm S$ (decay/scattering) we require $0.9 \approx \omega_{L'}/\omega_{pe} \pm \omega_S/\omega_{pe}$, where $\omega_S$ is a frequency available to ion-sound waves and $\omega_{L'}$ is a frequency available to the backwards branch of the Langmuir / electrostatic waves. For simplicity we can simply note the requirement $\omega_{L'}/\omega_{pe} > 1$, and so we require the presence of ion-sound waves with frequencies $\omega_S/\omega_{pe} > 0.1$, which is beyond the cut-off frequency ($\omega_{pi}/\omega_{pe} \approx 0.02$). Thus, for Run 2, such interactions are strongly suppressed as most of the beam mode occupies prohibited frequencies. On the other hand, for Run 1 the majority of the forward propagating waves are concentrated on the Langmuir dispersion curve and so they can interact with the low-frequency ion-sound waves, resulting in the development of the broad spectrum of counter-propagating waves as per (2), which go on to coalesce into harmonic emission as per (4). Our demonstration of this *decay inhibition* due to breaking of frequency matching requirements (wave beat conditions) is the first verification of the arguments made from a theoretical perspective by Cairns (1989).

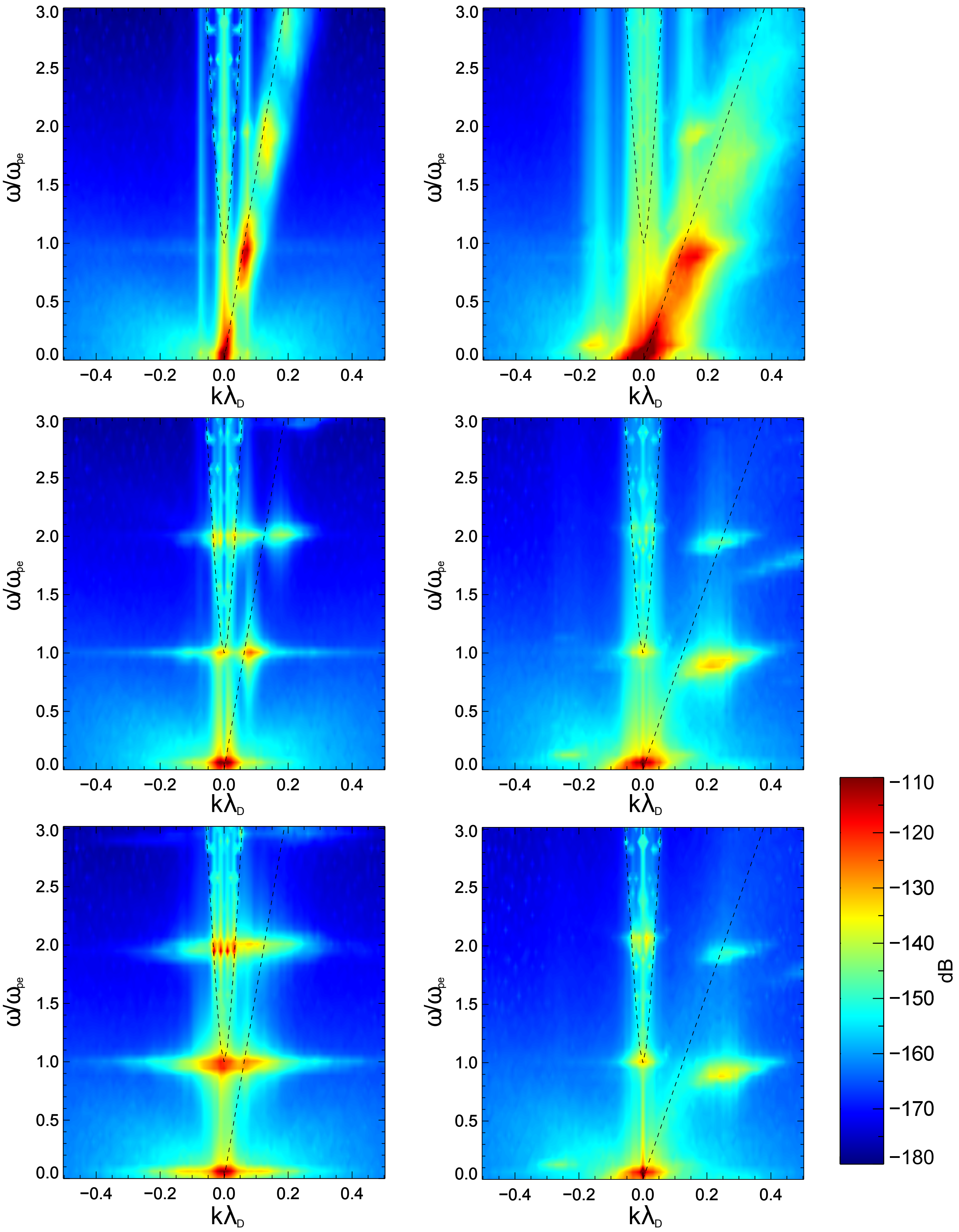

**Fig. 5.** Fast fourier transforms of $B_z$ for Run 1 (left) and Run 2 (right) over times $50 < t\omega_{pe}^{-1} < 150$, $450 < t\omega_{pe}^{-1} < 550$, $850 < t\omega_{pe}^{-1} < 950$, from top to bottom respectively.

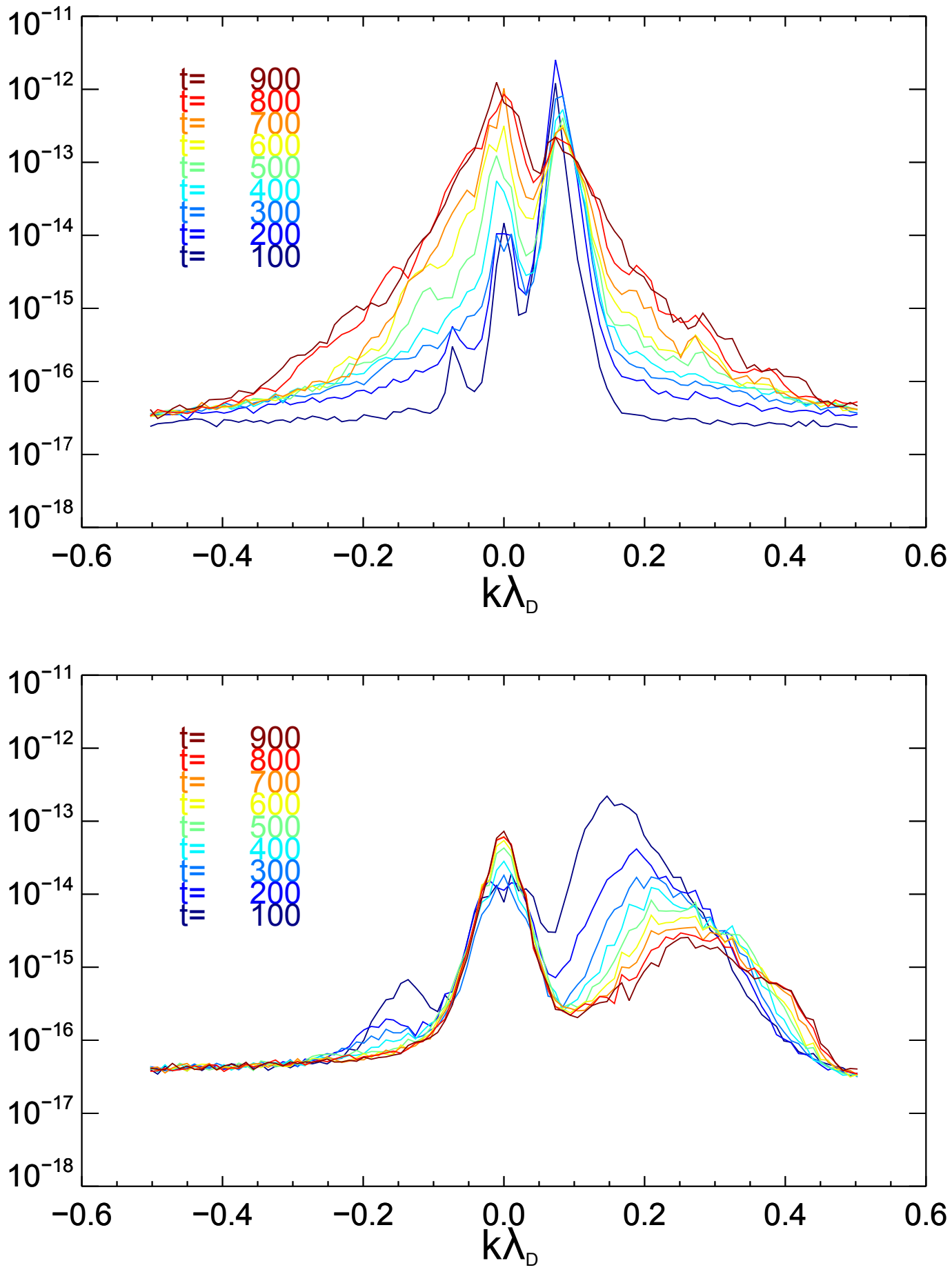

**Fig. 6.** Spectral energy density in $B_z$ ($\mathrm{J\,m^{-3}}$) at the fundamental frequency $\omega = \omega_{pe}$ for Run 1 (top) and Run 2 (bottom) for subsequent time windows.

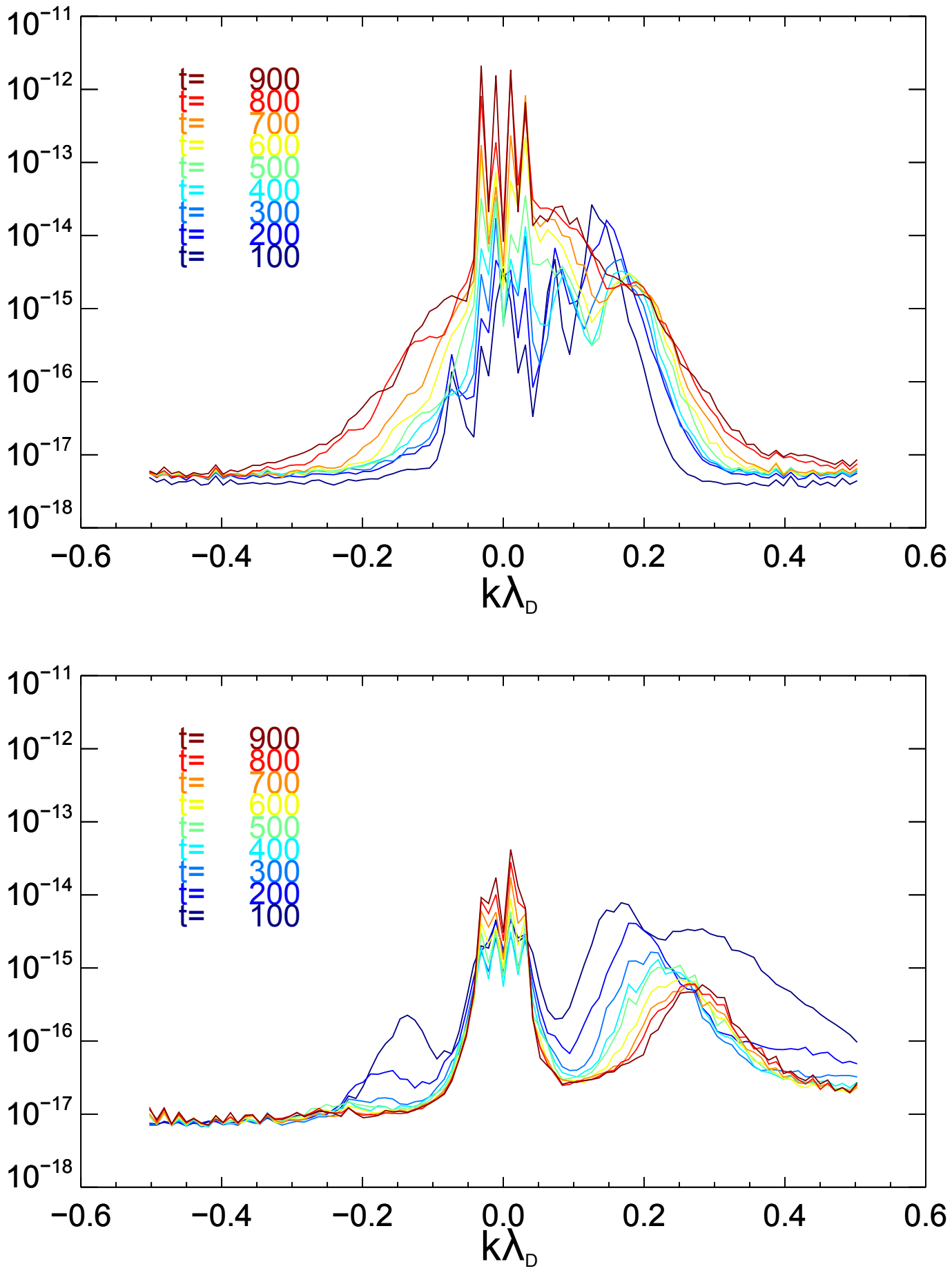

**Fig. 7.** Spectral energy density in $B_z$ ($\mathrm{J\,m^{-3}}$) at the harmonic frequency $\omega = 2\omega_{pe}$ for Run 1 (top) and Run 2 (bottom) for subsequent time windows.

The observed reduction in fundamental emission efficiency as per (3) can be explained similarly. Fundamental emission processes based on three-wave interactions of the type $L \pm S \rightarrow T_1$ demand frequency be conserved such that $\omega_L \pm \omega_S = \omega_{T_1}$. As the beam mode lies significantly below the $\omega_{pe}$, electromagnetic waves must be above it, the frequency conservation requirements require the low-frequency wave to be in excess of the cut-off, and thus they cannot be met.

In summary, we have presented two experiments for apparently comparable beams (at least in terms of the reactive/kinetic $P$ parameter, and a difference in $n_b/n_0$ of only one order of magnitude), whereby only in the case of Run 1 do we see significant and unambiguous fundamental and harmonic plasma emission far in excess of background levels. We propose that the demonstrated sensitivity, particularly to beam density, is the underlying explanation as to why Ganse et al. (2012b) have reported being able to find no evidence consistent with plasma emission in PIC simulations. Furthermore, we caution that as previous studies such as Kasaba et al. (2001); Umeda (2010) have interpreted their results as signs of emission in similar parameter regimes to Run 2 but not reported on noise thresholds, it is unclear if such systems actually generate efficient plasma emission that is compatible with observed radio emission. Indeed, typical setups (with dense beams) are closer to those used in Run 2 (i.e., the case where we find limited evidence for fundamental emission, and no evidence for harmonic emission.)

Ganse et al. (2012b), who could not demonstrate three-wave based emission from a single beam, explored different mechanisms to produce (harmonic) plasma emission. One notable mechanism, based on the action of two counter-propagating beams, was discussed in a series of papers (Ganse et al. 2012b,a, 2014). To illustrate some key differences in emission arising from the (single-beam) fundamental and harmonic plasma emission mechanisms and the two-beam emission mechanism, we reran Run 2 and included an additional, but otherwise identical counter-propagating electron beam (Run 3).

From early times we see strong signals in $B_z$ that are consistent with harmonic emission (Fig. 9). This is explained in that the beams directly drive electrostatic modes that match the beat conditions for harmonic emission (Fig. 10). As such, the generation of harmonic emission is guaranteed if the two beams have identical properties (as this determines their $\omega$ and $|k|$). It is unclear how robust this mechanism is in the case of non-identical beams of different $n_b/n_0$, $v_b/\Delta v_b$, $v_b/v_e$, and would make for interesting future work. We stress that this is not the 'classical' harmonic emission which relies on three subsequent stages (growth of forward propagating electrostatic waves, backscatter and decay to produce a counter propagating population, and their coalescence), and is rather directly powered by the artificial initial conditions.

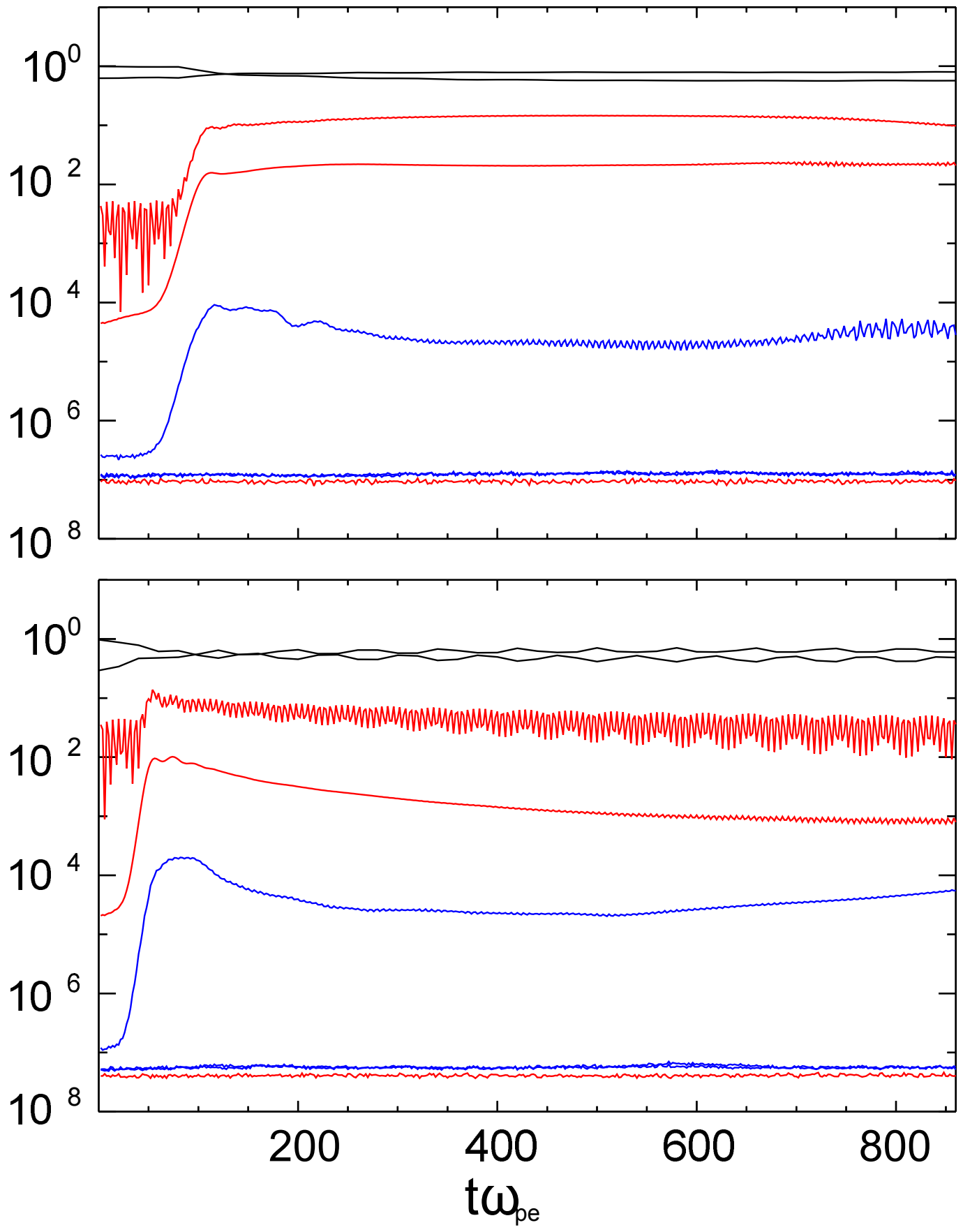

**Fig. 8.** Energies normalised to initial beam kinetic energy for Run 1 (top) and Run 2 (bottom). The black lines show the kinetic energy associated with the beam and background electrons (in the $x$-direction), the red lines the electric field energy (from $x$, $y$, and $z$ components when reading from top to bottom) and blue the magnetic field energy ($B_z$ component at $\sim 10^{-4}$, and $B_x$ and $B_y$ components at noise level $\sim 10^{-7}$).

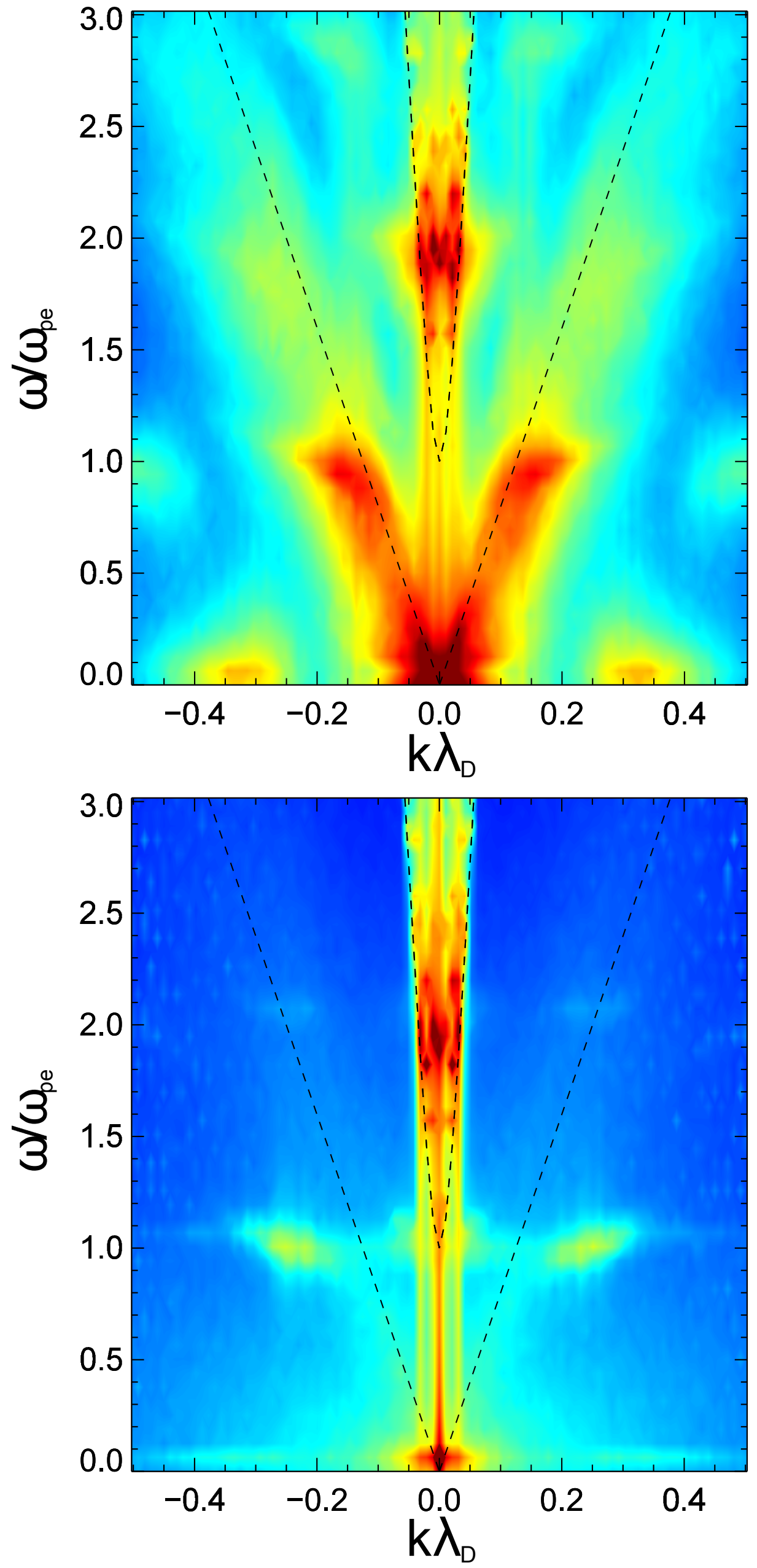

**Fig. 9.** Fourier transforms of $B_z$ for Run 3 at early (top, $50 < t\omega_{pe}^{-1} < 150$) and late (bottom, $850 < t\omega_{pe}^{-1} < 950$) times.

We also note that, compared to Run 2, we find little evidence for enhanced fundamental emission owing to the presence of the additional beam. This is consistent with the argument presented as to why $L \rightarrow S + T_1$ does not proceed efficiently Run 2; we observe in Fig 10 that the two beam modes are generated at the similar $\omega$ and $k$ as in Run 2 and so we do not expect a three wave interaction which is prohibited in Run 2 to be permitted due to the additional beam. Finally, we comment on the low-frequency enhancements in the electrostatic power spectrum. These are not ion-sound waves as their frequency is too high, being above the cut-off frequency. They are rather a direct, nonlinear plasma response to the ponderomotive force of the two crossing beam modes. Apparent harmonics of these 'daughter waves' (which are actually driven by the corresponding harmonics of the beam mode), can also be observed in Fig 10.

## 5. Conclusion

We have presented two numerical experiments for different beam-plasma systems which demonstrate a remarkable sensitivity in terms of resulting wave dynamics, with drastic consequences for plasma radio emission.

In Run 1 (a more tenuous and fast beam of $n_b/n_0 = 0.0057$, $v_b/\Delta v_b = v_b/V_e = 16$) , we find evidence consistent with all stages of the three-wave based fundamental and harmonic emission mechanisms, including the beam-mode to Langmuir mode coupling, the growth of a population of counter-propagating Langmuir/electrostatic waves via backscattering and decay processes, the action of fundamental emission, and the coalescence of the counter-propagating population to produce harmonic emission. For the first time using a fully kinetic and electromagnetic PIC simulation, we can confirm the role of all such stages whilst taking care to distinguish signals above the inherent noise levels associated with particle methods, and performing appropriate convergence testing. Thus, Run 1 is arguably the first unambiguous confirmation of the three-wave based emis-sion processes resulting from a single electron beam in the literature using the fully kinetic PIC approach.

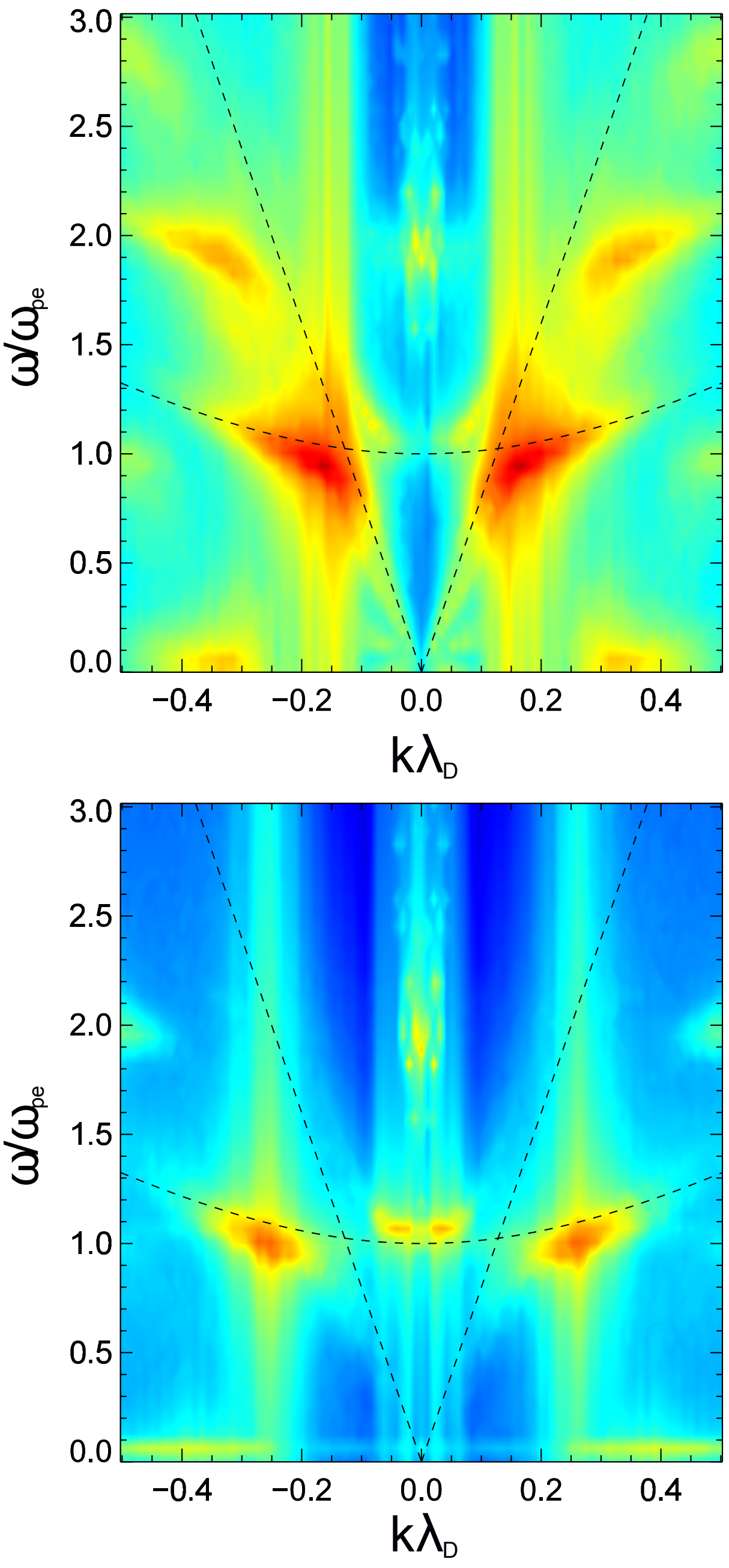

**Fig. 10.** Fourier transforms of $E_x$ for Run 3 at early (top, $50 < t\omega_{pe}^{-1} < 150$) and late (bottom, $850 < t\omega_{pe}^{-1} < 950$) times.

For Run 2 (a slower, denser beam of $n_b/n_0 = 0.05$, $v_b/\Delta v_b = v_b/V_e = 8$), we demonstrate the sensitivity of the parameter space by considering a more dense beam with $n_b/n_0$, a similar density to past works, and found that the processes are significantly suppressed due to the resulting non-Langmuir characteristics of the beam mode. Whilst a full parameter study may prove useful, but is beyond available computational resources, we hope that Run 2 demonstrates plainly that caution must be applied when attempting to simulate astrophysical beam-plasma systems using unrealistically dense beams. Whilst a larger density ratio reduces relaxation times (i.e. computer time), the results are unlikely to be physically representative of the intended system due to the sensitivity to beam parameters.

We also make the first connection between the action of the Weibel instability and the generation of an electromagnetic beam mode in the context of plasma emission. As this provides a stronger contribution to electromagnetic energy than the emission, we stress that evidence of emission in simulations must disentangle the two contributions (such as by our spectral approach) and not simply interpret changes in total electromagnetic energy as emission. Following Karlický (2009), who found that only very strong fields ($\omega_{pe}/\omega_{ce} \sim 1$) could inhibit this transverse behaviour, we expect that effect this is of importance to application to solar radio bursts (where the field is relatively weak).

Finally, comparison of our results indicate that, contrary to the suggestions of authors including Ganse et al. (2012b,a, 2014), the two beam, or counter-propagating beam, mechanism is not necessary to produce harmonic emission, and that in certain parameter spaces (such as Run 1), the single-beam emission mechanisms can proceed. However, that is not to say that the counter-propagating beam mechanism does not work. Rather, the suitability of either process depends on the physical situation and crucially, whether two beams are expected. In cases where we expect two counter-propagating beams to exist which also happen to be suitably connected to the Langmuir mode (primarily, this implies lower density ratios than those considered by (Ganse et al. 2012b,a, 2014), we anticipate competition between the two different mechanisms.

*Acknowledgements.* The authors acknowledge funding from the Leverhulme Trust Research Project Grant RPG–311. The computational work for this paper was carried out on the joint BIS, STFC and SFC (SRIF) funded DiRAC-1 cluster at the University of St Andrews (Scotland, UK). The EPOCH code used in this research was developed under UK Engineering and Physics Sciences Research Council grants EP/G054940/1, EP/G055165/1 and EP/G056803/1.